\documentclass[prc,superscriptaddress,showpacs,amssymb,amsmath,amsfonts,aps]{revtex4}

\usepackage{graphicx}
\usepackage{dcolumn}
\usepackage{bm}
\usepackage{epsfig}
\def\g1p{$G_1^p$}
\def\g1d{$G_1^d$}

\begin{document}

\title{Empirical Fit to electron-nucleus scattering.}

\newcommand*{\JLAB}{Thomas Jefferson National Accelerator Facility, Newport News, Virginia 23606}
\affiliation{\JLAB}
\newcommand*{\VIRGINIA}{University of Virginia, Charlottesville, Virginia 22901}
\affiliation{\VIRGINIA}
\newcommand*{\WM}{College of William and Mary, Williamsburg, Virginia 23187-8795}
\affiliation{\WM}

\author{P.E.~Bosted}
     \email{bosted@jlab.org}
\affiliation{\JLAB}
\affiliation{\WM}
\author{V. Mamyan}
\affiliation{\VIRGINIA}

\date{\today}


\begin{abstract}
An empirical fit to electron-nucleus scattering for $A>2$
is made based on world data. It is valid for $0<W<3.2$ GeV
and $0.2<Q^2<5$ GeV$^2$, and can be used with caution
at lower $Q^2$. The fit is based on previous
empirical fits to electron-proton and electron-neutron
scattering, taking into account the effects of Fermi
motion plus a substantial extra contribution that
fills in the dip between the quasi-elastic peak
and the $\Delta(1232)$ resonance.
\end{abstract}

\maketitle

\section{Introduction}
There are many applications in nuclear/particle physics
for a reliable parametrization of inclusive electron-nucleus
scattering. One example is the reliable evaluation
of radiative corrections to measured data to extract
inclusive electron scattering cross sections. Another
is the evaluation of background in electron scattering
from polarized ammonia. In this case, the contributions
from the unpolarized nitrogen relative to the polarized
proton must be taken into account to extract 
spin-dependent inclusive cross section asymmetries.

A prime motivation for a new empirical fit is the
availability of a large body of new, high precision
electron-nucleus scattering data from 
Jefferson Lab~\cite{Mamyan}.
The present fit was used in the evaluation of the
radiative corrections for much of these data.

Our basic fit form is similar to that of a previous
fit to inclusive electron-deuteron scattering~\cite{Bosted}.

\section{Definitions and Kinematics}

In terms of the incident electron energy, $E$, the 
scattered electron energy, $E^{'}$, and the scattering angle, 
$\theta$, the absolute value of the exchanged 4-momentum squared 
in electron-nucleon scattering is given by
\begin{equation}
Q^2 = (-q)^2 =  4EE^{'}{\sin}^2 \frac{\theta}{2}, 
\end{equation}
and the mass of the undetected hadronic system is
\begin{equation}
W^2 = M_p^2 + 2M_p\nu -Q^2,  
\end{equation}
with $M_p$ the proton mass, $\nu = E-E^{\prime}$, and the small
terms involving the electron mass squared have been neglected. 

In the one-photon exchange approximation, the spin-independent 
cross section for inclusive electron-nucleon scattering can 
be expressed in terms of the photon helicity coupling as
\begin{equation}
\frac{d\sigma}{d\Omega dE^{'}} = 
\frac{\alpha_F^2 \cos^2(\theta/2) } 
     {[2 E  \sin^2(\theta/2)]^2} (W_2(W^2,Q^2) + 2 \tan^2(\theta/2) W_1(W^2,Q^2)]
\label{eq:cs1}
\end{equation}
where $\alpha_F=1/137$ is the fine structure constant, and the structure
functions $W_1$ and $W_2$ are each the sum of a quasi-elastic 
($W^{QE}$) and
an inelastic $(W^I)$ piece, as detailed below. 
The quasi-elastic fit is given in terms of the $F_1$ and $F_2$ structure
functions, which are related to $W_1$ and $W_2$ by
$F_1 = M W_1$ and $F_2=\nu W_2$.

\section{Inelastic Fit function}
The functional form of the fit is:
$$W_1^I = (W_1^F(W^2,Q^2) + W_1^{MEC}(W^2,Q^2)) 
f_{EMC}(x^\prime)$$
where $W_1^F$ is a Fermi-smeared sum of free proton
and neutron contributions, $W_1^{MEC}$ is an
additional term to fill in the dip between quasi-elastic
and $\Delta(1232)$ peaks, and $f_{EMC}$ is an
unpublished parametrization of the ``EMC'' effect
(nuclear dependence of Deep Inelastic Scattering)
by S. Rock in 1994.

To obtain $W_2^I$, we use the relation:
$$W_2^I = W_1^I  [1 + R_A(W,Q^2)] / (1 + \nu^2 / Q^2)$$
where 
$$R_A(W,Q^2) = R_p(W,Q^2) (1 + P_6 + P_{23} A)$$
where $R_p$ is the ratio of longitudinal to transverse
cross sections from the fit to world proton data
given in Ref.~\cite{Christy}. The fit parameters $P_i$ are listed
in Table~\ref{tab:param}.

\subsection{Inelastic term $W_1^F(W^2,Q^2)$}
To obtain the inelastic term, $W_1^F(W^2,Q^2)$,
we use the equation:
\begin{equation}
W_1^F(W^2,Q^2) = C(x)\sum_i [
Z W_1^p((W_i^\prime)^2,Q^2) +
(A-Z) W_1^n((W_i^\prime)^2,Q^2)] f_i
\end{equation}
where the correction term depending on $x$ Bjorken
is given by:
\begin{equation}
C(x) = 1 + P_{13}x + P_{14}x^2 + P_{15}x^3 + 
      P_{16}x^4 + P_{17}x^5,
\end{equation}
and $W_1^p$ and $W_1^n$ are the free proton~\cite{Christy}
and neutron~\cite{Bosted} structure functions. 
The shifted values $W_i^\prime$ are given by
\begin{equation}
(W_i^\prime)^2 = W^2 + \xi_i k_F |\vec q|
- 2 E_s (\nu + M)
\end{equation}
\noindent where $|\vec q|^2 = Q^2 + \nu^2$, 
\begin{equation}
\xi_i = -3 + 6(i-1)/98
\end{equation}
\noindent and 
\begin{equation}
f_i = 0.0245  e^{(-\xi_i^2/2)}.
\end{equation}
\noindent The sum is nothing more than a step-wise integration
over a Gaussian whose width is controlled by a Fermi
momentum $k_F$, truncated at $\pm 3\sigma$, with
a shift in central $W$ related to the binding energy
$E_s$. The values of $k_F$ and $E_s$ used for
the different nuclei are given in Table~\ref{tab:kfes}.

\begin{table}[ht]
\centering
\begin{tabular}{ccccccccc}
\hline
$A$ & $k_F$ (GeV) & $E_s$ (GeV) \\
3 & 0.115 & 0.001 \\
$3<A<8$ & 0.190 & 0.017 \\
$7<A<17$ & 0.228 & 0.0165 \\
$16<A<26$ & 0.230 & 0.023 \\
$25<A<39$ & 0.236 & 0.018 \\
$38<A<56$ & 0.241 & 0.028 \\
$55<A<61$ & 0.241 & 0.023 \\
$A>60$ & 0.245 & 0.018 \\
\hline
\hline
\end{tabular}
\caption{Values of Fermi-broadening parameter $k_F$
and binding energy parameter $E_s$ for different nuclei.}
\label{tab:kfes}
\end{table}

\subsection{``MEC'' term}
To fill in the dip between the quasi-elastic and 
$\Delta(1232)$ resonance peaks, we added an extra
term, which we dubbed the ``MEC'' (meson-exchange
current) term. The importance of this term grows
with $A$. The form of $W_1^{MEC}$ is:
\begin{equation}
W_1^{MEC} = \frac{P_0}{f} e^{
-[(\sqrt{W^2} - P_1)^2]/P_2]}
\end{equation}
where
\begin{equation}
f=(1 + (Q^2)^\prime) / P_3 )^{P_4} \hskip .15in  \nu^{P_5}
\hskip .15in ( 1 + P_{18}  A^{(1 + P_{19} x)} )
\end{equation}
where $(Q^2)^\prime)$ is the larger of 0.3 GeV$^2$
or $Q^2$ and $x=Q^2/2M\nu$. 

\subsection{$f_{EMC}$}
\begin{figure}[hbt]
\centerline{\includegraphics[height=4.0in]{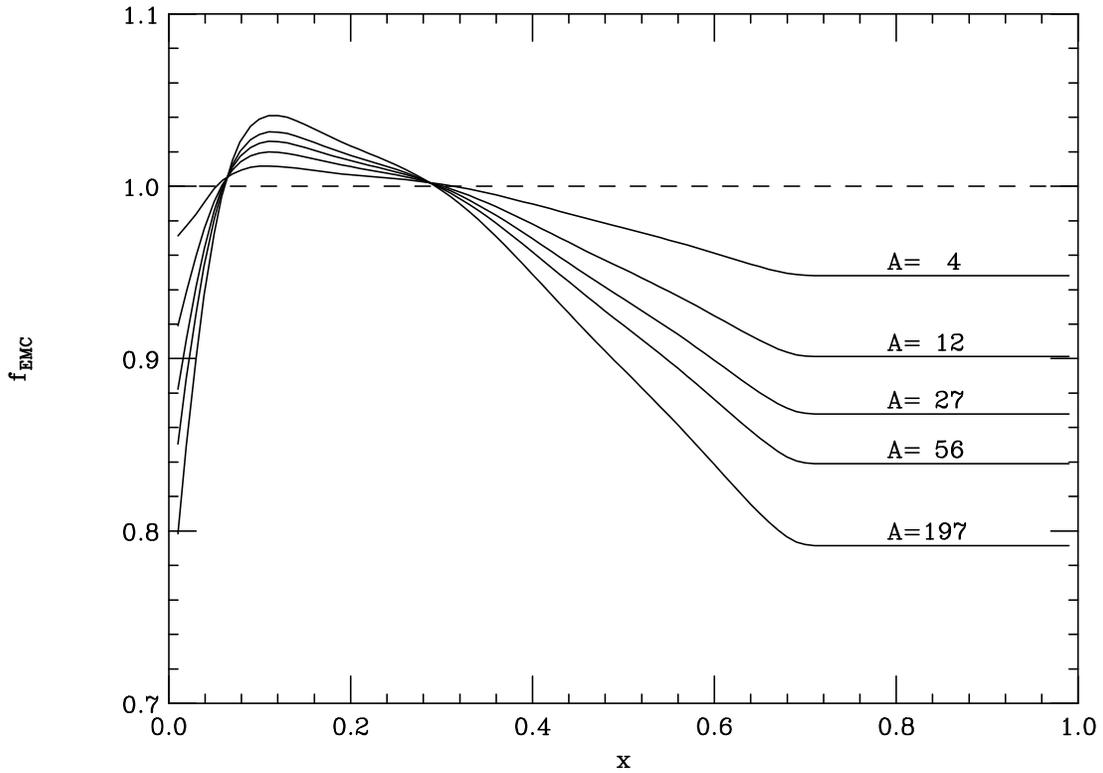}}
\caption{Illustration of the $x$-dependence of the
function $f_{EMC}$ for five values of atomic number A.}
\label{fig:emc}
\end{figure}

The function $f_{EMC}$ is  given by:
$$f_{EMC} = c A^\alpha$$
where 
$$\alpha = \alpha_0 + \sum_{i=1}^8 \alpha_i (x^\prime)^i$$
where $x^\prime$ is the smaller of 0.7 and $Q^2/2M\nu$
and the coefficients $\alpha_i$ are given in 
Table~\ref{tab:alpha}, and where
$$c=e^{0.0169 + 0.01809 x^\prime + 0.0050427 (x^\prime)^2}$$
The $f_{EMC}$ fit is illustrated for five representative
nuclei in Fig.~\ref{fig:emc}. Note that the original
fit was only for data with Bjorken $x<0.7$, that is why
we ``freeze'' the results above that value

\begin{table}[ht]
\centering
\begin{tabular}{ccccccccc}
\hline
$\alpha_0$ & 
$\alpha_1$ & 
$\alpha_2$ & 
$\alpha_3$ & 
$\alpha_4$ & 
$\alpha_5$ & 
$\alpha_6$ & 
$\alpha_7$ & 
$\alpha_8$ \\
       -0.069887 &
        2.1888 & 
       -24.667 & 
        145.29  &
       -497.23 & 
        1013.1 & 
       -1208.3 &
        775.76 & 
       -205.87 \\
\hline
\end{tabular}
\caption{Values of the $\alpha_i$ EMC fit parameters.}
\label{tab:alpha}
\end{table}

\section{Fit parameters}
The fit parameters are given in Table~\ref{tab:param},
except for $P_{18}$. This parameter, which is important
in the strength of the MEC term, was fit individually
for $^3He$, $^4He$, C, Al, and Cu. We assume nearby
nuclei will have the same parameter values. The results
are given in Table~\ref{tab:p18}.

\begin{table}[ht]
\centering
\begin{tabular}{llrr}
\hline
$P_{ 0}$ &     0.005138 & $P_{ 1}$ &     0.980710 \\ 
$P_{ 2}$ &     0.046379 & $P_{ 3}$ &     1.643300 \\ 
$P_{ 4}$ &     6.982600 & $P_{ 5}$ &    -0.226550 \\ 
$P_{ 6}$ &     0.110950 & $P_{ 7}$ &     0.027945 \\ 
$P_{ 8}$ &     0.406430 & $P_{ 9}$ &     1.607600 \\ 
$P_{10}$ &    -7.546000 & $P_{11}$ &     4.441800 \\ 
$P_{12}$ &    -0.374640 & $P_{13}$ &     0.104140 \\ 
$P_{14}$ &    -0.268520 & $P_{15}$ &     0.966530 \\ 
$P_{16}$ &    -1.905500 & $P_{17}$ &     0.989650 \\ 
$P_{18}$ &   see Table 2 & $P_{19}$ &    -0.045536 \\ 
$P_{20}$ &     0.249020 & $P_{21}$ &    -0.137280 \\ 
$P_{22}$ &    29.201000 & $P_{23}$ &     0.004928 \\ 
\hline
\end{tabular}
\caption{Values of the fit parameters.}
\label{tab:param}
\end{table}

\begin{table}[ht]
\centering
\begin{tabular}{ll}
\hline
A & $P_{18}$ \\
3 & 70 \\
4 & 170 \\
$4<A<21$ & 215 \\
$20<A<51$ & 235 \\
$A>50$ & 230 \\
\hline
\end{tabular}
\caption{Values of $P_{18}$ for different nuclei.}
\label{tab:p18}
\end{table}

\section{Quasi-elastic contribution}
The quasi-elastic contribution is calculated using
the equations in Ref.~\cite{Sick}. The free nucleon
factors are taken from Ref.~\cite{FF}. These form
factors are based on inclusive electron scattering,
and thus have the 2-photon corrections appropriate
to the present quasi-elastic fit. The values of
Fermi motion parameter $k_F$ and binding energy $E_s$
are slightly different than in Ref.~\cite{Sick}: we
used those in Table~\ref{tab:kfes}. We used a 
multiplicative Pauli suppression factor given by
$$(3/4) (|\vec q| / k_F)(1 - (|\vec q|/ k_F)^2)/12)$$
for $|\vec q| < 2k_F$, otherwise no correction was made.
We assumed the same suppression for $W_1$ and $W_2$.
For the scaling function, we used~\cite{Fy}
\begin{equation}
F(\psi^\prime) = 1.5576 / k_F
(1 + 1.7720^2 (\psi^\prime + 0.3014)^2) 
(1 + e^{-2.4291 \psi^\prime})
\end{equation}
We used the same function for both $W_1$ and $W_2$.

The nominal results from the above were corrected
as follows:
$$F_1=F_1^{nom}
(1 + P_7 + P_8 y + P_9 y^2 + P_{10} y^3 + P_{11} y^4)$$
and
$$ R = R^{nom} (1 + P_{12} )$$
\noindent where $R^{nom}$ is the ratio of longitudinal
to transverse cross sections, defined using:
$$R^{nom} = (F_2^{nom} / \nu) (M / F_1^{nom}) 
(1. + \nu^2 / Q^2) - 1$$
and where $y=(W^2 - M^2) / |\vec q|$. The fit parameters
$P_7$ to $P_{12}$ are given in Table~\ref{tab:param}.

\section{Coulomb Corrections}
Coulomb corrections are taken into account in the
simple energy gain/loss method, using a slightly
higher incident and scattered electron energies
at the vertex than measured in the lab.
The shifts are the same in both cases, and are
given by~\cite{Coulomb}:
\begin{equation} 
     V  = 0.775 \hskip .05in \frac{3}{2} \hskip .05in \alpha_F (Z-1)/R_0
\end{equation}
\noindent where $R_0$ (in units of GeV) is given by:
\begin{equation}
      R_0     = 1.1 A^{(1/3)} + 0.86 A^{(-1/3)}
\end{equation}

\section{Discussion}
\begin{figure}[hbt]
\centerline{\includegraphics[height=2.3in]{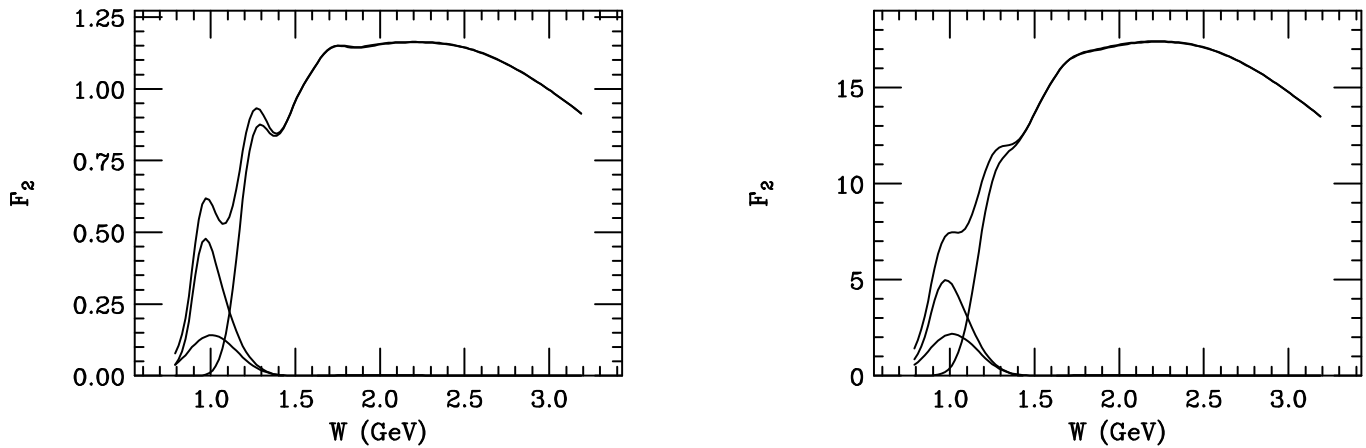}}
\caption{Illustration of the $W$-dependence of the
structure function $F_2(W,Q^2)$ for He (left) and Fe
(right), for $Q^2=0.6$ GeV$^2$. The upper curves are
the sum of the three lower curves (quasi-elastic,
``MEC'', and inelastic, from left to right).}
\label{fig:f2}
\end{figure}

To illustrate the main features of the fit, the
response function $F_2$ is plotted versus $W$ 
for He (left) and Fe (right) in Fig.~\ref{fig:f2}.
The top curve in each plot is the sum of the
there components: a quasi-elastic peak centered
on $M$, a smaller but broader ``MEC'' peak centered
near $W=1.05$ GeV, and the inelastic continuum.
The main two features that can be noticed are that
the quasi-elastic and $\Delta(1232)$ peaks are
wider in Fe than in He (due to the larger Fermi
momentum), and the MEC contribution is relatively
larger in Fe than in He.

\begin{figure}[hbt]
\centerline{\includegraphics[height=5.5in]{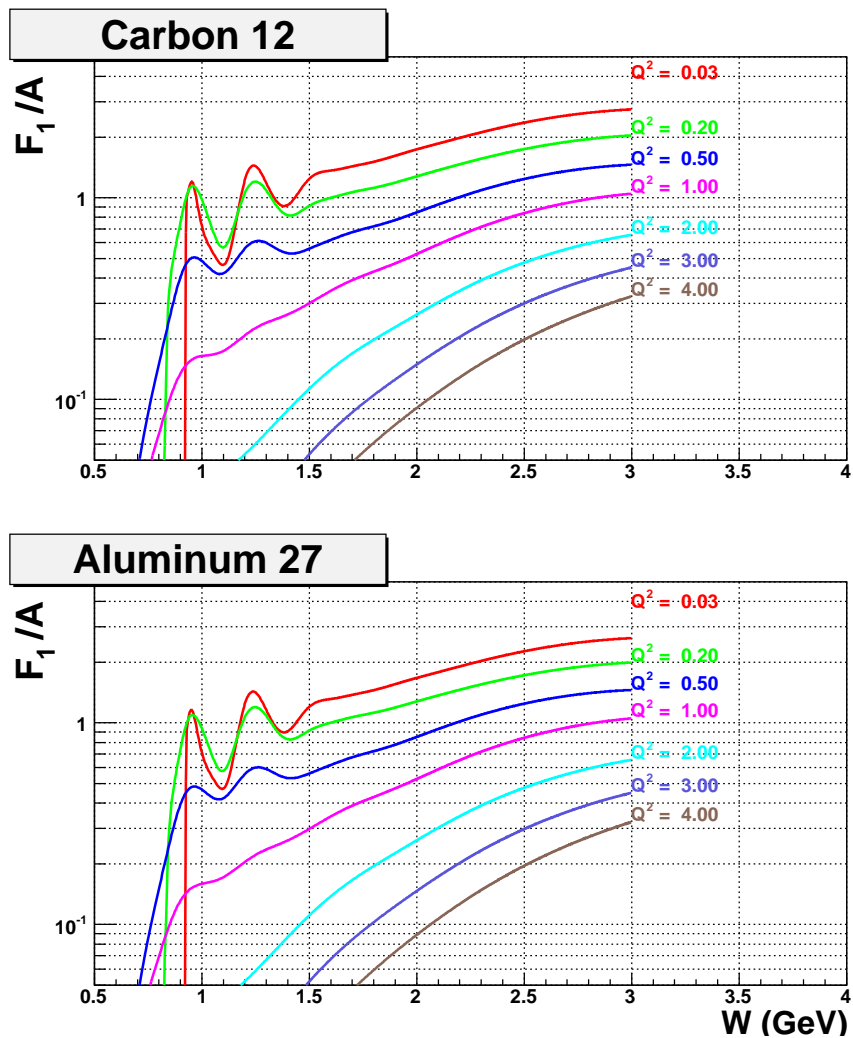}}
\caption{Illustration of the $W$-dependence of the
structure function $F_1(W,Q^2)$ (per nucleon) 
for C and Al for a wide range of $Q^2$ values (units
are GeV$^2$).}
\label{fig:vahe2}
\end{figure}

The $Q^2$ dependence of $F_1$ is illustrated for C and
Al in Fig.~\ref{fig:vahe2}. Note the very prominent
quasi-elastic and $\Delta(1231)$ peaks at low $Q^2$,
which ``disappear'' rapidly at higher $Q^2$. This feature
is what makes at fit accurate to better than 10\% so
difficult for $Q^2<0.2$ GeV$^2$.

\begin{figure}[hbt]
\centerline{\includegraphics[height=5.5in]{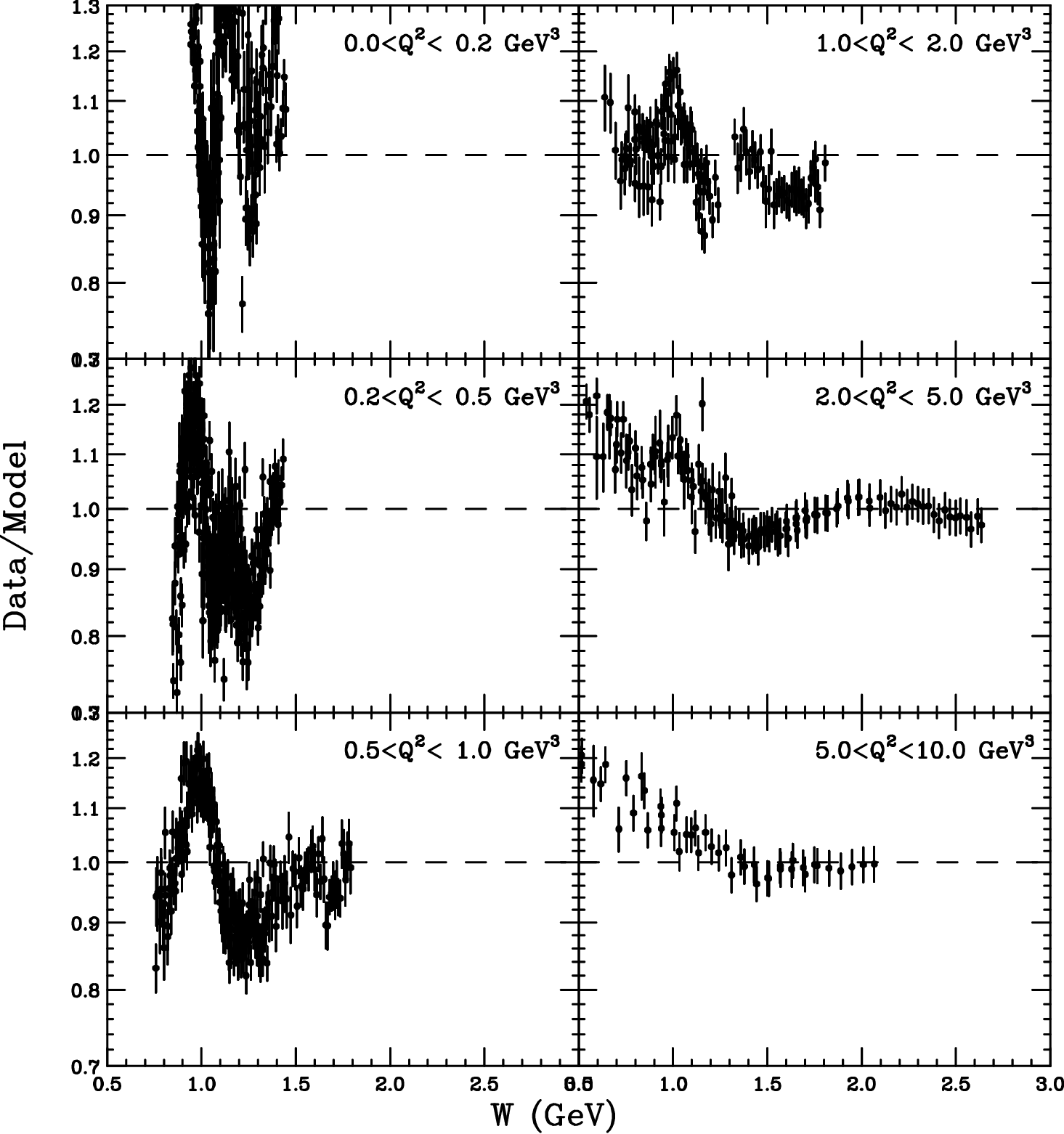}}
\caption{Ratios between fit and world data for $^4$He in 
six $Q^2$ bins as a function of $W$.}
\label{fig:He}
\end{figure}

\begin{figure}[hbt]
\centerline{\includegraphics[height=5.5in]{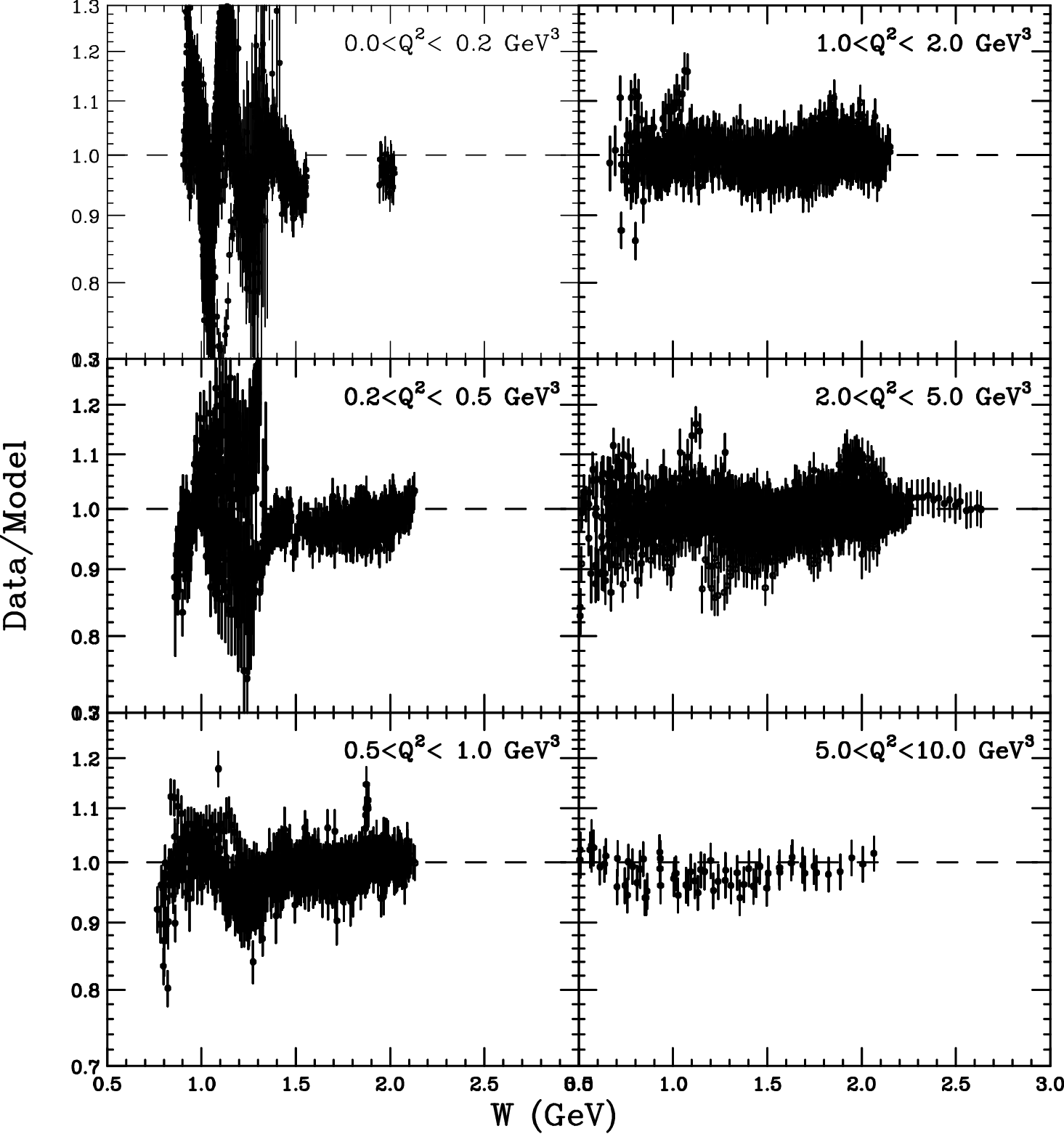}}
\caption{Ratios between fit and world data for C in 
six $Q^2$ bins as a function of $W$.}
\label{fig:C}
\end{figure}

\begin{figure}[hbt]
\centerline{\includegraphics[height=5.5in]{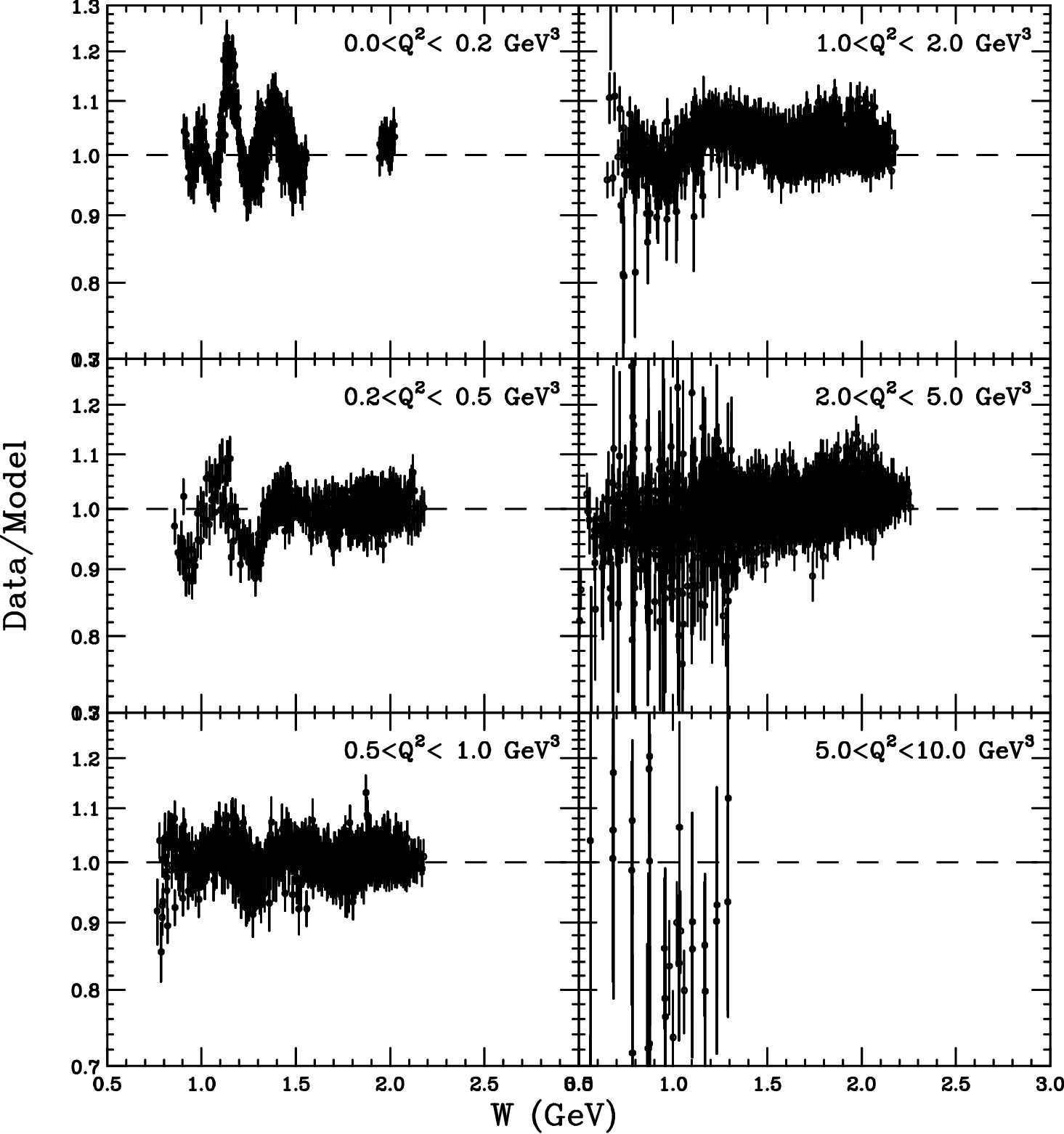}}
\caption{Ratios between fit and world data for Al in 
six $Q^2$ bins as a function of $W$.}
\label{fig:Al}
\end{figure}

\begin{figure}[hbt]
\centerline{\includegraphics[height=5.5in]{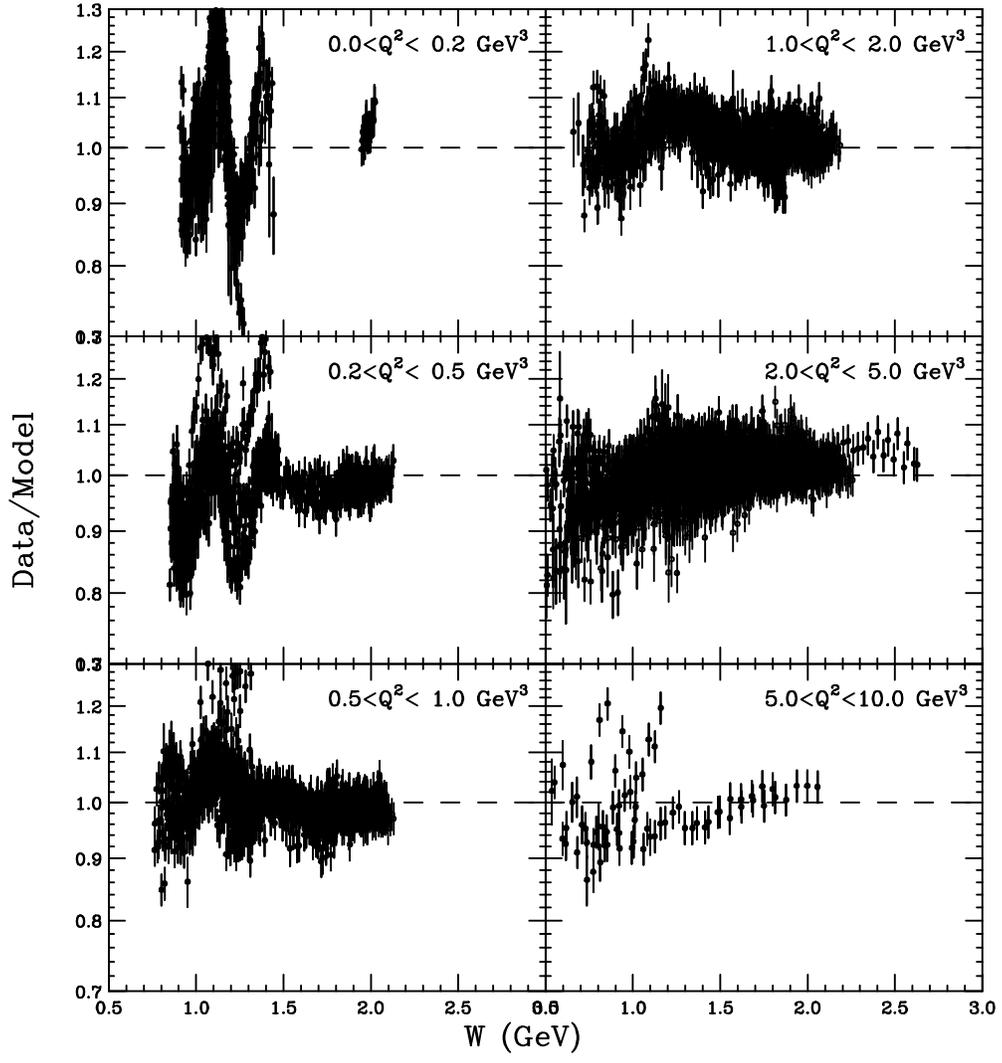}}
\caption{Ratios between fit and world data for Fe and Cu in 
six $Q^2$ bins as a function of $W$.}
\label{fig:Fe}
\end{figure}

\begin{figure}[hbt]
\centerline{\includegraphics[height=2.3in]{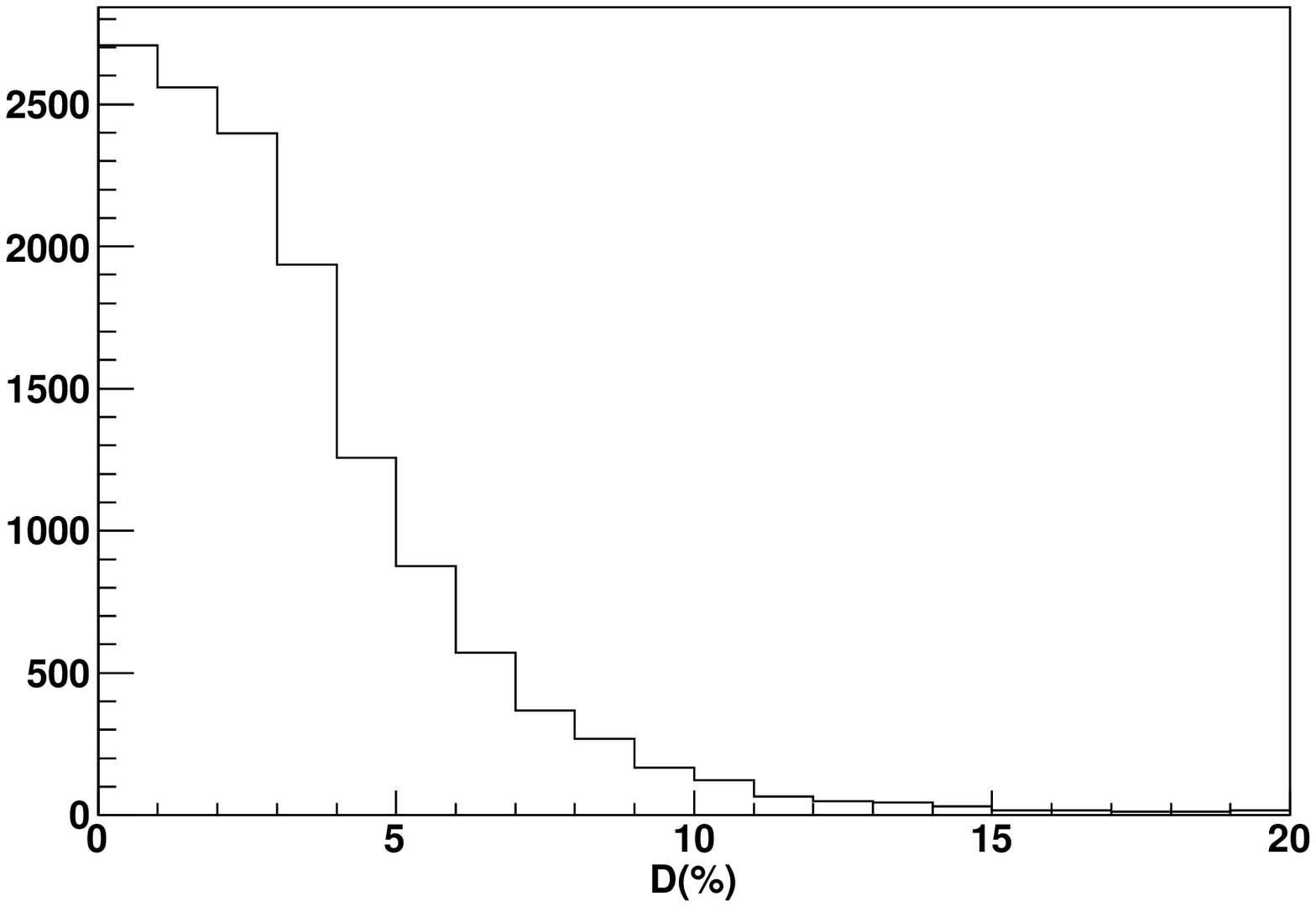}}
\caption{Frequency distribution of deviations between
data and fit.}
\label{fig:vahe3}
\end{figure}

The fit is compared with world data~\cite{Day,Mamyan}
 for He, C, Al, and Fe/Cu
in Figs.~\ref{fig:He}-\ref{fig:Fe}. For $W>1.2$ GeV, the
agreement is generally within 5\%, and 
better than 3\% on average.
In the quasi-elastic peak region, 
some larger oscillations around
unity can be seen due to the difficulty 
in matching the precise
shape of the quasi-elastic peak with 
actual data, but on average
the agreement is within 5\%. The biggest 
discrepancies are seen
at low $Q^2$, where the limitations of the 
plane-wave impulse
approximation for quasi-elastic scattering 
become most apparent.
Figure~\ref{fig:vahe3} shows the frequency distribution
of the deviations (in percent) between data and fit
for all the data points used.

\section{Source code}
The stand alone FORTRAN source code 
(named F1F209.f) for this fit
is available~\cite{Fit}.

\section{Acknowledgments}
The authors are grateful for useful discussions
with J.E. Amaro, J. Arrignton, M.B. Barbaro, M. Christy, D. Day, 
W.T. Donnely, and D. Gaskell.

\end{document}